\documentstyle[aps,preprint,12pt]{revtex}
\begin{document}
\draft
\title{DYNAMICS OF INTERACTING CLUSTERS AND DIELECTRIC RESPONSE  IN RELAXOR FERROELECTRICS}

\author{B.E.Vugmeister and H. Rabitz }
\address {\it Department of Chemistry, Princeton University, Princeton, NJ
 08544}

\maketitle

\begin{abstract}
The dielectric response in relaxor ferroelectrics is analyzed in the
framework a  model for the polarization
dynamics in the presence of polar clusters.
We associate the origin of polar clusters  with the
atoms displaced from their centro-symmerical positions even above $T_c$.
Their collective hopping in multi-well
potentials induced by disorder is analogous with the situation in
glasses. The theory explicitly takes into account the distribution of
cluster reorientation frequencies  and the effect of
cluster-cluster interactions in highly polarizable crystals, which we 
describe in terms of  the  local field distribution function.
The dielectric constant is obtained from an  integral master
equation for the polarization dynamics in the presence of a time
dependent electric field. The  theory is applied for the
analysis of the shape of the frequency dependent permittivity in the 
typical relaxor ferrolectrics $PST$ as a
function of temperature. The comparison of the theory with experiment
shows that in contrast to earlier assumptions, 
 the observed Vogel-Fulcher dependence of the permittivity
maximum is a consequence of the Vogel-Fulcher temperature dependence
of the cluster reorientation frequency.

\end{abstract}
\draft

~~~~~~~~~~~~~~~

\section{Introduction}

In the family of ferroelectrics a large  group of mixed   and disordered
 "relaxors" has been identified  including perovskites 
 $PbMg_{1/3}Nb_{2/3}O_3$  (PMN), 
$PbSc_{1/2}Ta_{1/2}O_3$  (PST),  La-modified  $ PbZr_{1-x}Ti_x O_3$ (PLZT), tungsten-bronze structure oxides like $Pb_x Ba_{1-x}Nb_2O_6$ (PBN), etc.\cite{Smolensky,Cross,Burns,Vieland,Mathan,Kleemann,Setter,Sommer,Egami,VugT,Vakhrushev,Samara,Glazounov,Qian,Khachaturian}
Due to the effect of configurational disorder the properties of relaxor 
ferroelectrics are very different from those in translationally invariant
ferroelectrics.
Among the unusual properties of these  materials are a) the
coexistence of slow kinetics typical for spin glasses with a very large
dielectric constant indicating intermideate range polar order on the
nanometer length scale which can be transformed  to true long range order,
({\it i.e.}, to macroscopic polarization and strain) by a suitable change of the
composition or by applying an external electric field;
b) the existence of a frequency dependent slim
hysteresis loop even above the transition temperature; c) 
the difference between field-cooled and zero field-cooled dielectric 
susceptibility, etc. 
All these findings
indicate that the observed properties of relaxor ferroelectrics are
nonequilibrium properties.  Relaxor ferroelectrics thus 
represent a new low temperature state of polar dielectrics compared
with conventional ferroelectrics. 

Experiments show that there is  symmetry breaking  on a nanometer 
scale as observed from Raman scattering\cite{Burns} and X-ray and neutron
diffraction\cite{Mathan,Egami}. 
This observation implies that polar clusters exist 
even well above $T_c$ and the observed properties of relaxors are
strongly affected by the reorientations of the clusters. In particular,
cluster reorientation in the applied electric field induces strong
polar-strain coupling which makes relaxor ferroelectrics the candidate 
materials for the next generation of ultrasonic transducers\cite{Shrout}.

It has been proposed\cite{Cross}, in analogy with the cluster model of spin 
glasses, that the polar clusters behave like large superparaelectric dipole 
moments. The broad distribution of relaxation times for cluster orientations 
originates from the distribution of the potential barriers separating 
different orientational states. The superparaelectric model\cite{Cross} 
based on the assumption of independent clusters can not explain, however, 
the appearance of long range order with the change of the material 
composition or due to an applied external field. 
The appearance of long range order also can not be explained by the inclusion 
of frustrated interactions between superparaelectric moments\cite{Vieland}
that lead to the dipole spin glass state.
An alternative proposition\cite{Kleemann,Glazounov,Qian,Khachaturian} relates  the origin of
relaxor  behavior  to  the domain states induced by the static random 
field caused, for example, by charged composition fluctuations. 
Although this model seems very attractive, it encounters difficulties
in explaining the observed cluster dynamics in the high temperature phase.

 We will show that the
proposed model is capable to describe the anomalies of dielectric response of 
relaxor ferroelectrics and can be applied for  systems possessing  first or second order phase 
transitions or remaining only incipient ferroelectrics with very high
dielectric constant and vanishing spontaneous polarization.
 
The origin of polar clusters in relaxor ferroelectrics is still not
understood. It has been proposed\cite{VugT} that off-center ions might be 
responsible for relaxor properties in analogy with 
$K_{1-x}Li_xTaO_3$ (KLT) where $Li$ impurity ions occupy  
the off-center positions near the vacant $K$ sites.  
It is well established\cite{VugG} that  in the case of impurity induced relaxors  like KLT  the
single impurity potential possesses multi-well structure which allows the
thermal jumps of off-center ions  between different potential minima.  
In contrast, in disordered complex perovskites 
the displacement of atoms from their centro-symmetrical positions  
is caused apparently by the charge compositional fluctuations, 
which violate the 
charge balance within the adjacent unit cells and lead to the additional
electrostatic forces on the atoms.
In this situation one would not expect that the single atom
potential energy possesses  multi-minima structure.  
Atom reorientations would more probably take
place if they associate with  the  collective motion of atoms within 
small clusters, ({\it i.e.}, the cluster  potential energy is
characterized by the multi-well (double well) structure).

Such a picture is analogous to  the situation  in glasses where the origin  of
two level systems responsible for glass anomalies is associated with 
the double well potentials\cite{AHV,Phillips}. It is generally
assumed that the existence of the double     
well potentials is due to the disorder in glasses, so that  local
rearrangements of atoms might switch the system between two adjacent local
energy minima.  Attempts to detect 
double well potentials in computer simulations\cite{Weber} based on the assumption of the motion of 
single atoms\cite{Harris} have been unsuccessful. Instead the origin of double well potentials in
glasses has been successfully explained\cite{Silbey} in terms of collective
motion of atoms within small clusters.  

In this paper we develop further the cluster model of relaxor
ferroelectrics taking explicitly into account a) the broad
distribution of local field experienced by each cluster due to
cluster-cluster interaction and b) the broad distribution of potential
barriers controlling cluster dynamics.
We will show that the
proposed model is capable of describing the anomalies of dielectric response of
relaxor ferroelectrics and can be applied for  systems possessing  first or second order phase 
transitions or remaining an incipient ferroelectrics with very high
dielectric constant and vanishing spontaneous polarization.

\section{Model}

Based on the analogy between disorder induced double well potentials
in glasses and disordered relaxor ferroelectrics 
we will adopt the following picture of the cluster dynamics in 
relaxor ferroelectrics.
Each minimum in a double well potential for a given cluster is
characterized by the cluster dipole moment or cluster polarization 
$P_{cl}$. 
The potential barrier between different minima
determines the atom hopping frequency $\tau^{-1}$. Clusters interact
with each other, and interacting reorientable polarizable clusters 
should increase  the crystal dielectric response.
This conclusion is supported by the experiment\cite{Setter} in PST where the effect of
disorder in the relative occupation of B sites by Sc or Ta atoms dramatically
increases  the dielectric constant.

Time or frequency dependent dielectric response can be analyzed with 
the use of the  master equation
\begin{equation}
{d P_{cl} \over d t}=-{1 \over \tau}(P_{cl}-P_{cl}^{eq}(E_L)),
\label{eq:a7}
\end{equation}
well known from the theory
of ferroelectrics of order-disorder type\cite{Blinc,Lines}, describing the
relaxation of the polarization of each cluster to its quasi-equilibrium
value $P_{cl}^{eq}(E_L)$ which depends on the value of the local field
$E_L$ induced by other clusters at any moment of time.   
In general the local field $E_L$  is a time 
dependent random field. It also  includes   the contribution from
the applied field $E_{ex}$ and the contribution from the static random
fields caused by the  material imperfections.

In order to apply Eq.(\ref{eq:a7}) to relaxor ferroelectrics  one should perform the average
over the distribution of relaxation times $\tau$ and the distribution
of the local fields $E_L$.
For this purpose we rewrite 
Eq.(\ref{eq:a7}) in the equivalent integral form 
and take the average with respect to $\tau, E_L$ and the initial cluster 
polarization $P_{cl}(0)$ to obtain
\begin{equation}
P(t)=P(0) \tilde{Q}(t) -\int_0^t d t'{\partial \tilde{Q}(t-t') \over \partial t} P^{eq}(t').
\label{eq:a9}
\end{equation}

\noindent
The function $\tilde{Q}(t)=\overline{e^{-t/\tau}}$, where the overbar denotes 
the average over the relaxation time $\tau$,   
characterizes the slow nonexponential kinetics of the system. In addition 

\begin{equation}
P^{eq}(t)=\int d E  P_{cl}^{eq}(E) f(E,P(t)), 
\label{eq:a11}
\end{equation}

\noindent
where $f(E, P(t))$ is the distribution function of the local field
which  depends 
parametrically on the value of the average  polarization of the system $P(t)$. 
We will consider below $f(E, P)$ in the form    
$f(E, P) = \tilde{f}(E-\gamma P -\gamma \epsilon_0 E_{ex}/ 4 \pi)$.
This form of $f$
is consistent with the mean field approximation $f(E, P)=\delta (E - \gamma P-\gamma \epsilon_0 E_{ex}/4 \pi )$,  
where $\delta$ is the delta-function and  $ \gamma $  is the
local field phenomenological parameter. 
In disordered  systems the effect of composition fluctuations leads to a deviation from 
the simple mean field picture that can be taken into account by the replacement of the 
$\delta$-function  by the function $\tilde{f}$ with  finite 
width. The shape and the width of $\tilde f(E)$ depend on the explicit
form of the 
cluster-cluster interactions  as well as on  static random fields caused by 
the material imperfections.

The value $\gamma  \epsilon_0 E_{ex}/4 \pi $  is the local field induced by 
the external field in the dielectric media with the dielectric constant 
$\epsilon_0 >>1$ \cite{Kittel}
({\it i.e.,} we assume that the polar clusters are distributed in a highly 
polarizable dielectric media). 
In  relaxor ferroelectrics which are mainly perovskite-based highly polarizable materials,
the typical values are $\epsilon_0 > 10^3$. 

Eq.(\ref{eq:a9}) can be applied for the analysis of 
different experimental 
situations ({\it e.g.,} decay of the polarization, the difference 
between field-cooled and zero 
field-cooled dielectric susceptibility, the effect of a frequency  dependent 
hysteresis loop, etc). 
In this paper we will concentrate, for illustration,  on the
calculation of the frequency dependent linear dielectric susceptibility.
In order to calculate the steady state  susceptibility 
in the presence of an additional time independent field $E_{ex}^{(0)}$ we 
write
$E_{ex}(t) = E_{ex}^{(0)} +  E_{ex}^{(1)}e^{i \omega t}$ and 
$P(t) =P_s +P_1(t)$,
where $P_s$ is the time independent polarization induced by the field 
$E_{ex}^0$ (for $E_{ex}^{(0)} =0$, we denote $P_s$ as the spontaneous
polarization). We obtain from Eq.(\ref{eq:a9}) the following 
 self consistent equation for $P_s$

\begin{equation}
P_s =\int d E P_{cl}^{eq}(E) f(E,P_s).
\label{eq:a17}
\end{equation}
 Assuming $P(0) = P_s $  in Eq.(\ref{eq:a9}), 
considering  a linear expansion of $P^{eq}(t)$ with 
respect to $E_{ex}^{(1)}$ and $P_1(t)$, and taking Laplace transform
of Eq.(\ref{eq:a9}), we obtain

\begin{equation}
\epsilon (\omega, T) = {\epsilon_0  \over 1-  \kappa(T) Q(\omega, T)},
\label{eq:a19}
\end{equation}
where we used the definition of the dielectric constant $\epsilon(\omega,T) = 
4 \pi \partial P_1(\omega)/ \partial E_{ex}^{(1)} +\epsilon_0$. 
In Eq.(\ref{eq:a19})

\begin{equation}
\kappa(T) =\int d E  P_{cl}^{eq}(E)  {\partial  f(E, P_s)\over \partial P_s},
\label{eq:a20}
\end{equation}

\noindent
and $Q(\omega, T)$ is the Laplace transform of $ \partial \tilde {Q(t)} 
/ \partial t $  given by 

\begin{equation}
Q(\omega,T)=\overline {(1/(1 +i \omega \tau))}
\label{eq:a3}
\end{equation} 
Assuming an Arrhenius or Vogel-Fulcher (VF) law for $\tau$ 

\begin{equation}
\tau(T) = \tau_0 \exp{[U /(T-T_0)]},
\label{eq:a4} 
\end{equation}
we may calculate $Q(\omega,T)$. Thus, we obtain  for $Q'(\omega,T)$\cite{Courtens}

\begin{equation}
Q'(\omega, T)= \int_0^{(T-T_0)\ln{1 \over \omega \tau_0}} dU g(U),
\label{eq:a5}
\end{equation}
where g(U) is the distribution function of the potential barriers.

In the spirit of Landau phenomenological theory one can  further
expand the right hand side of Eqs.(\ref{eq:a17})
and (\ref{eq:a20}) in a power series with respect
to $P_s$ assuming that the applied field is sufficiently small, and
present $\kappa$ as 
 
\begin{equation}
\kappa = a_1 +3 a_3 P_s^2 +5 a_5 P_s^4
\label{eq:a25}
\end{equation}
Eqs.(\ref{eq:a19}), (\ref{eq:a5}), and   (\ref{eq:a25}) give the 
phenomenological description of the dynamical response of 
relaxor ferroelectrics and can be applied for  systems possessing  first or second order phase 
transitions or remaining as only incipient ferroelectrics with very high
dielectric constant and vanishing spontaneous polarization.

\section{Amomalies in the dielectric response}

In this section we illustrate a
capability of the model in the description of the dielectric response
of relaxor ferroelectrics.

{\it  Relaxors with incipient ferroelectric order}.
In order to simultaneously reproduce  the high values of  the dielectric constant in  relaxor ferroelectrics and the absence of spontaneous polarization 
(like that in PMN or PST with vacancies) one should assume that  $a_3
< 0$ and $a_1 (T) \rightarrow 1$ remaining, however, less than 1 at
all temperatures.  
We chose for the illustrative calculations $a_1(T) = 0.95 \tanh({ 0.5 / T})$
(which, according to Eq.(\ref{eq:a19}), reproduces Curie-like high
temperature behavior of  $\epsilon(0,T)$ and its high saturation value at low 
temperatures) and $g(U)= 2.5 U^4/(0.5 +U^5)^2$.
We assume also the Arrhenius temperature dependence of $\tau(T)$,
i.e. $T_0 =0$ in Eq.(\ref{eq:a4}). Using the chosen values of
$a_1(T), g(U)$, and $T_0$ we calculated the dielectric permittivity $\epsilon$ 
as a function of  temperature for 
different values of $\ln({1\over \omega \tau_0})$.

The calculated real part of permittivity is shown in Fig.1
Note  that the frequency dependent maximum of $\epsilon'$
is not just a  relaxation maximum. It
originates from the competitive temperature dependences of  $Q'(T)$
and  $a_1(T)$.  The behavior obtained for  $\epsilon'$ is in  reasonable
qualitative agreement with the experiment\cite{Kleemann,Setter,VugT}. 
For a more detailed comparison with the experiment one needs to find
the functions $Q(\omega,T)$ and $a_1(T)$ corresponding to the best fit with the
experimental data.

{\it Manifestation of the first order phase transition}.
In some relaxor ferroelectrics (e.g., disordered PST or KLT above the
critical concentration of Li ions) 
 there is  evidence of a first order phase 
transition, which manifests itself in the sharp drop of the dielectric constant at the temperature below the position of the relaxation maximum.
In order to reproduce this behavior within the proposed
phenomenological theory one should assume $ a_3 > 0$. 
For example for  $a_3=1.7,  a_5=-11$ and the values of  $a_1(T)$ and  $Q'$ 
being the same  as used above, the first order phase transition occurs
at $T_c \approx 0.2 $ (in the chosen dimensionless units).

The calculated values of $\epsilon'$  are shown in Fig.2.  
A remarkable feature of the $\epsilon'$ temperature dependence  is that
 the  relaxation maximum of $\epsilon'$ 
approaches the 
phase transition temperature with a decrease of the frequency 
until it finally disappears transforming to a sharp peak like that in
conventional ferroelectrics. This behavior is in qualitative agreement
with that observed in PST.

{\it Reconstruction of the relaxation function $Q(\omega,T)$}.
PST with B-site chemical disorder  undergoes a first order 
relaxor-ferroelectric phase transition at $T_c \approx 269 K$\cite{Setter}. 
The dielectric permittivity of PST shows pronounced frequency
dispersion with the 
position of a frequency dependent maximum obeying the VF law\cite{Setter}
$\omega =\omega_0 \exp[-U/ (T_m-T_0)]$,
where $\omega$ is the frequency of the applied field and  $T_m$ is the 
temperature of the permittivity maximum corresponding to the frequency  $\omega$ .  
It has been widely accepted (see, e.g., \cite{Binder}) that the  VF type relation for the permittivity 
maximum  is a consequence of the VF law (\ref{eq:a4}) for $\tau(T)$. This assumption has been argued recently by Tagantsev
\cite{Tagantsev} who proposed that the observation in PST of the VF  frequency 
dependence of $T_m$ can be explained with the use of the  Arrhenius temperature dependence of $\tau$ taking into account 
the existence of the first order phase transition and the fact that in PST
$T_0 \approx T_c$. An important conclusion of Tagantsev's analysis is the  
indication that $T_m$ might be influenced significantly by the  values
of static permittivity, not only by the specific temperature 
dependence of $\tau$. However, in order to be more conclusive 
one should analyze the shape of the frequency dependent permittivity
as a function of temperature for 
different frequencies, not just the frequency dependence of the  
permittivity maximum. Such an analysis is given below with the use of 
Eq.(\ref{eq:a19}).

First, we extrapolated the experimental data\cite{Setter} and extracted the 
static permittivity above $T_c$ as shown in Fig.3 (dotted curve). 
Then, using  Eq.(\ref{eq:a19}) and the fact that at $\omega=0 ~~Q'$ is
equal to 1, we obtained the values of $\kappa(T)$ above $T_c$. We 
used the value $\epsilon_0 \approx 500 $ for the host lattice
permittivity based on the experimental 
low temperature values of $\epsilon(T)$ where the polar
clusters are frozen.
With the values obtained  for  $\kappa(T)$ and  the experimental values of the frequency dependent 
permittivity $\epsilon'(\omega_1,T)$ at frequency $\omega_1 =10KHz$ (curve 1)
we have calculated the values of $Q'(\omega_1,T)$ using
Eq.(\ref{eq:a19}). 
According to Eq.(\ref{eq:a5}) $Q'(\omega, T) $ is a function of 
$(T-T_0)\ln (\omega \tau_0)$ that results in the scaling relation 
$Q'(\omega, T) = Q'(\omega_1, T_1)$ with 
$T_1 = (T-T_0){\ln(\omega \tau_0) \over \ln(\omega_1 \tau_0)} +T_0 $.
Using this  relation one can reconstruct the values of $Q'(\omega, T)$ 
at other frequencies and, therefore, the values of $\epsilon'(\omega,T)$.  
The results of such a 
reconstruction are presented in Fig.3 for the upper 100Hz (curve 2)
and lower 1MHz (curve 3) boundary frequencies used in the experiment.
We used  $\tau_0 \approx 10^{-12}$ in accordance with 
Ref.\cite{Setter}.

One can see from Fig.3 that  the  
approach  reproduces  rather well the shape of the temperature 
dependence of the permittivity at different frequencies of the applied field.
The parameter of the fit $T_0$ is found to be $T_0 \approx 258K$
which is in  reasonable 
agreement with the value $T_0\approx 265K$ estimated in Ref.\cite{Setter}.    
This
result means that the employment of the VF law for $\tau(T)$ is very
crucial for obtaining the 
correct temperature and frequency dependence of the permittivity for PST. 
The VF relation for $\tau(T)$
results simultaneously in the very fast temperature dependence of the  
permittivity and its  rather slow   dependence on
$\ln(\omega\tau_0)$, observed in the experiment. At the same time an 
Arrhenius like dependence of $\tau(T)$ would result in an extremely rapid dependence of  the
permittivity on $\ln(\omega \tau_0)$, which is inconsistent  with the 
experimental data.   

We are grateful to  A. Khachaturian, A. Tagantsev, and J. Toulouse for
the useful discussions. 
  This work is   supported  by  the National Science Foundation
and the Army Research Office.

\newpage

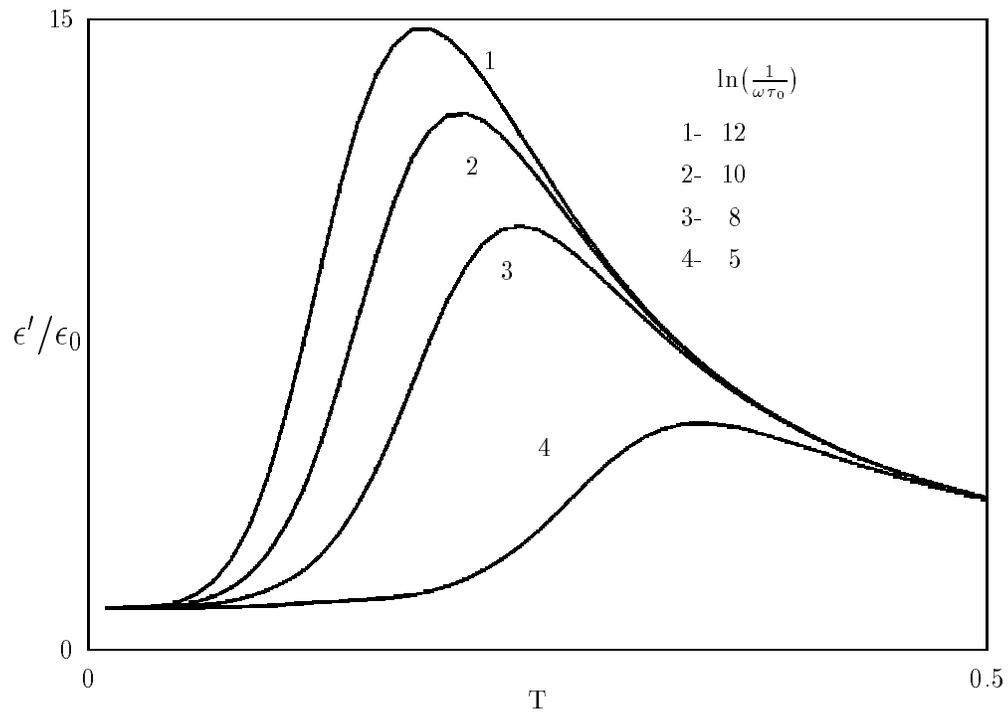
\begin{figure}
\caption{Real part of the dielectric susceptibility in relaxors with
incipient ferroelectric order.}
\end{figure}

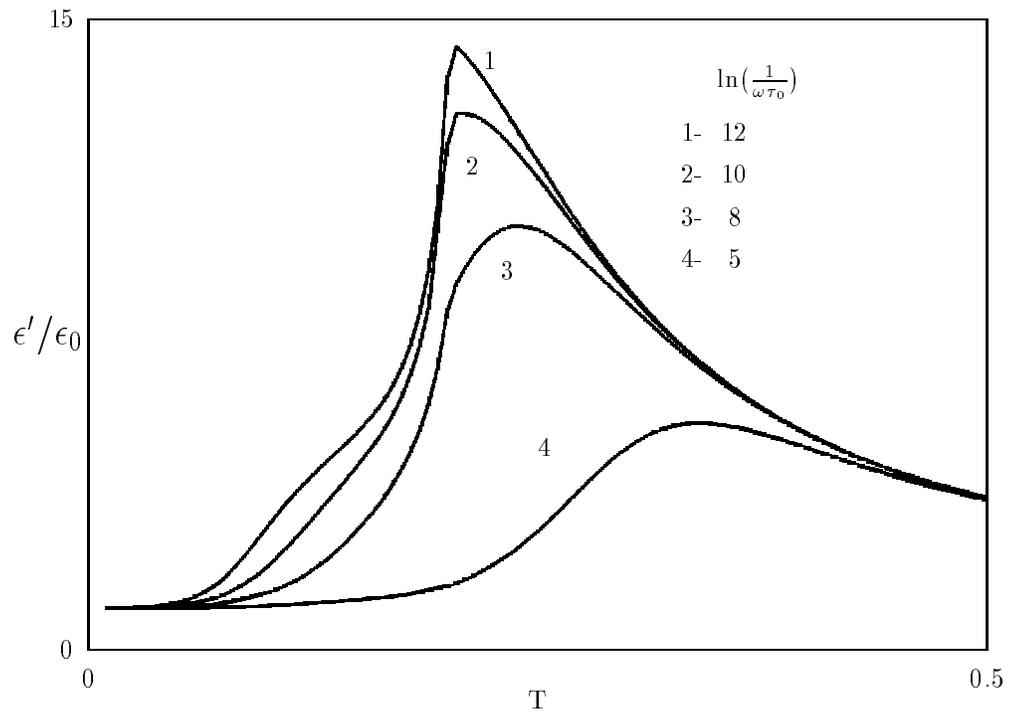
\begin{figure}
\caption{Effect of the first order phase transition on the dielectric
response in relaxor ferroelectrics.}  
\end{figure}

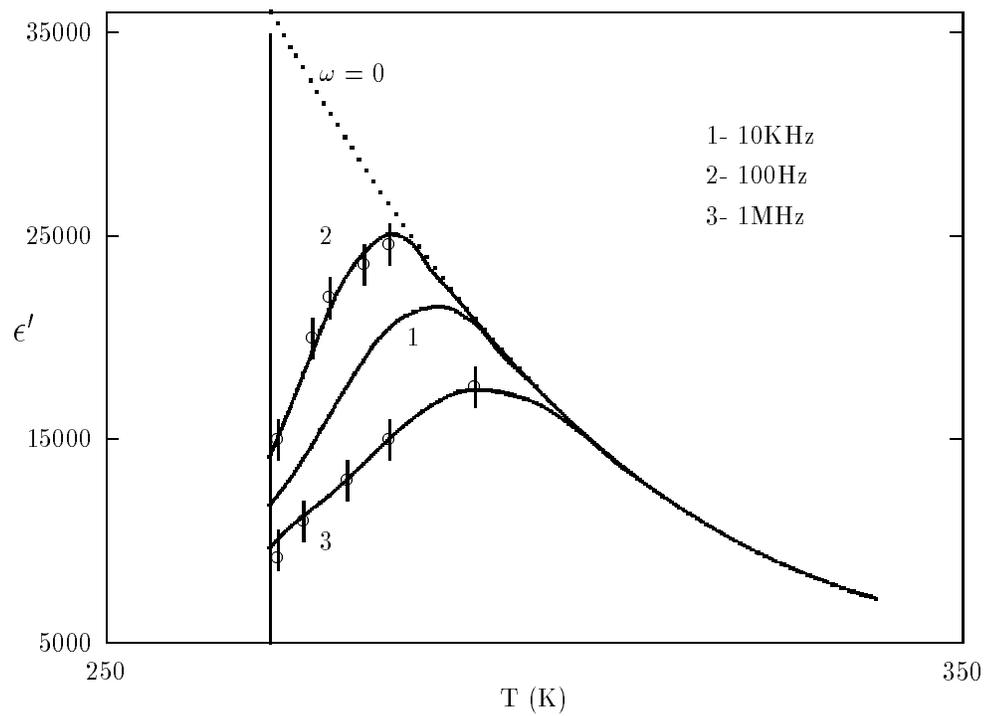
\begin{figure}
\caption{$\epsilon'$(T) for disordered PST above $T_c$ for different 
frequencies. The vertical  line shows the phase transition
temperature. Solid lines 2,3 and the circles are respectively the 
reconstructed and the experimental values of $\epsilon'(T)$. 
The error bars indicate the uncertainties of the reconstruction due to
the   uncertainties in the  values obtained for  $\kappa(T)$ by the
extrapolation(dotted line) of experimental data.}
\end{figure}

\end{document}